\documentclass{article}

\usepackage{spconf,amsmath,graphicx,multirow,float,amsmath,amssymb,mathrsfs}
\usepackage{booktabs} 
\usepackage{tabularx}
\usepackage{url}
\usepackage[colorlinks,
            linkcolor=red,
            anchorcolor=black,
            citecolor=black]{hyperref}
\usepackage{hyperref}
\title{MMCOSINE: MULTI-MODAL COSINE LOSS \\ TOWARDS BALANCED AUDIO-VISUAL FINE-GRAINED LEARNING}
\name{Ruize Xu\textsuperscript{1}, Ruoxuan Feng\textsuperscript{1}, Shi-Xiong Zhang\textsuperscript{2}, Di Hu\textsuperscript{1}\sthanks{Corresponding author. Di Hu is also a Research Fellow of Metaverse Research Center, Renmin University of China, Beijing, P.R.China, 100872.}}
\vspace{-2.5cm}
\address{\textsuperscript{1}Gaoling School of Artificial Intelligence, Renmin University of China, Beijing, China\\
\textsuperscript{2}Tencent AI Lab, Bellevue, WA, USA}

\begin{document}
%\ninept
%
\maketitle

\begin{abstract}

% Image-based fine-grained task has been widely discussed over the last decade of research, while audio-visual fine-grained task receives less attention, even it combines effective information from additional modalities and may have significant advantages than single visual modality. \textbf{Meanwhile, the issue of imbalanced multi-modal learning has also been studied recently towards a better joint-learning performance.} In our paper, \textbf{we first analyse the imbalance phenomenon and its potential cause in joint audio-visual fine-grained learning via observations on both feature distribution and the modulus of weight and feature vectors}. Further, to alleviate this imbalance problem, we propose a novel multi-modal cosine loss function that performs a modality-wise $L_2$ normalization to remove radial variation, based on which the final decision is determined by the combination of two cosine similarity scores. This loss also provides a simple but informative way to view the implicit constraints on the inter-modality discrepancy in concatenation-based a8/udio-visual learning. To verify our method, we present experiments on several audio-visual fine-grained tasks, including speaker verification, emotion recognition and fine-grained categorization. \textbf{Our new approach achieves better performance than baselines on all the tasks. We also have further discussions and experiments to prove that our method is highly versatile in multiple modality scenarios, applicable to the state-of-the-art fusion strategy and flexible with margin-based methods. }
Audio-visual learning helps to comprehensively understand the world by fusing practical information from multiple modalities. However, recent studies show that the imbalanced optimization of uni-modal encoders in a joint-learning model is a bottleneck to enhancing the model's performance. We further find that the up-to-date imbalance-mitigating methods fail on some audio-visual fine-grained tasks, which have a higher demand for distinguishable feature distribution. Fueled by the success of cosine loss that builds hyperspherical feature spaces and achieves lower intra-class angular variability, this paper proposes Multi-Modal Cosine loss, MMCosine. It performs a modality-wise $L_2$ normalization to features and weights towards balanced and better multi-modal fine-grained learning. We demonstrate that our method can alleviate the imbalanced optimization from the perspective of weight norm and fully exploit the discriminability of the cosine metric. Extensive experiments prove the effectiveness of our method and the versatility with advanced multi-modal fusion strategies and up-to-date imbalance-mitigating methods. The project page is \url{https://gewu-lab.github.io/MMCosine/}.
\vspace{-0.2cm}
\end{abstract}
%
% \begin{keywords}
% fine-grained, audio-visual, multi-modal, speaker verification, emotion recognition, imbalance
% \end{keywords}
%
\section{Introduction}
\vspace{-0.2cm}
\label{sec:intro}
% Fine-grained learning, the key challenge of which is to discriminate objects that are highly similar in overall appearance but differ in fine-grained features \cite{wei2021fine}, has received significant interest in the past years, with the learning objectives developing from static, specific, visual-only entities like cars, animals and clothes \cite{krause20133d,van2021benchmarking,liu2016deepfashion} to dynamic, abstract, multi-modal concepts like actions and emotions \cite{shao2020finegym,abdul-mageed-ungar-2017-emonet}. However, fine-grained learning that combines multi-modal information has been poorly explored especially in audio-visual field, while it's human nature to discriminate things from multiple modalities. For example, two species of birds may have similar appearance but distinguishable voice and vice versa. Although not directly mentioned, several audio-visual tasks can be classified as fine-grained learning due to their fine-grained features of each modality, such as speaker verification and emotion recognition. Meanwhile, the task of audio-visual fine-grained categorization has been studied with expanded image dataset \cite{zhang2018audio,van2022exploring}, laying the groundwork for the study of audio-visual fine-grained tasks.
Fine-grained learning, the critical challenge of which is the small inter-class and large intra-class variation of similar subordinate categories \cite{wei2021fine}, has received significant interest in the past years, with the learning objectives developing from entities \cite{van2021benchmarking} to concepts \cite{abdul-mageed-ungar-2017-emonet,shao2020finegym}. However, multi-modal fine-grained learning combining multi-sensory information has been poorly explored, especially the mechanism of how several modalities mutually learn and cooperate. 
% In uni-modal fine-grained learning task, a variety of margin-based cosine loss deriving from vanilla softmax bundled with cross-entropy have shown their power. Inspired by these mature methods, we propose a new loss function with modality-wise $L_2$ normalization for common mid-concatenation fusion and find it effective to alleviate the imbalance possibly caused by the dominant part of weight parameters. We evaluate our methods with three audio-visual tasks mentioned above: speaker verification, emotion recognition and bird categorization and find our method makes better result on most of the experiments, which reveals that our method makes multiple encoders jointly trained better. Also, we discuss the implicit information within the new framework and its expandability.

\begin{figure}[t]

  \flushright{\includegraphics[width=7.8cm]{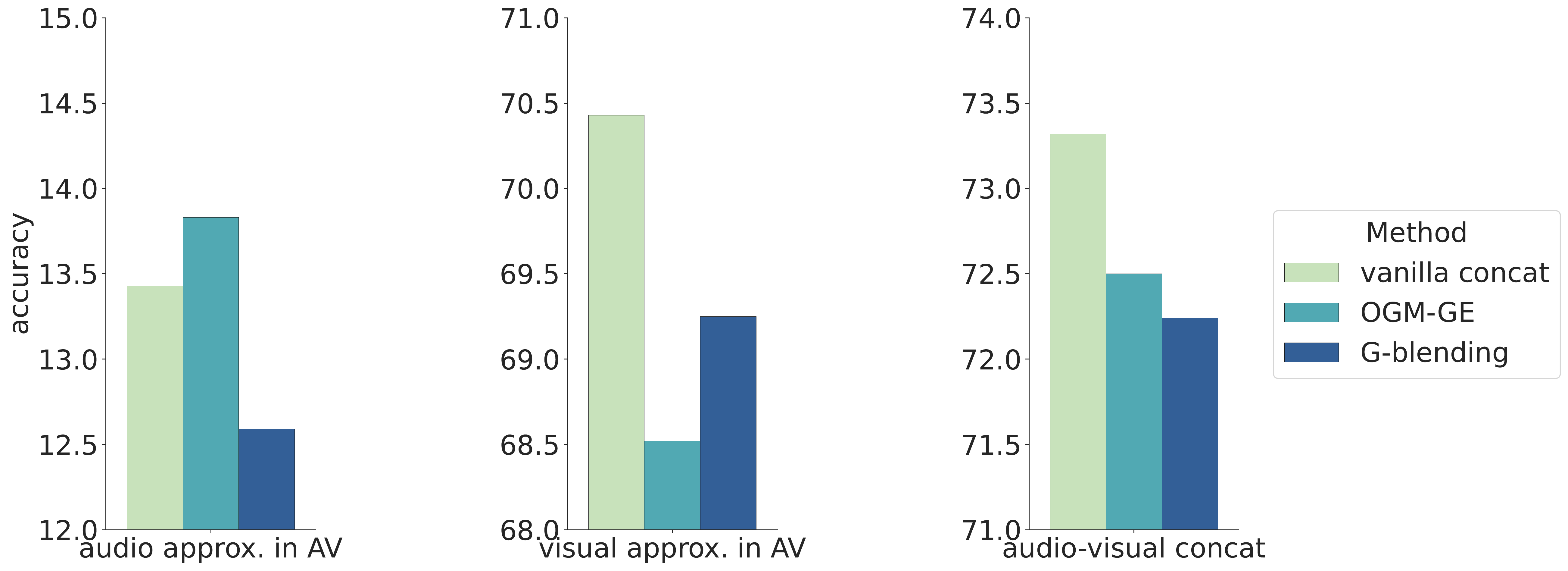}}
  \vspace{-0.4cm}
\caption{The change of the uni-modal and overall accuracy between the latest imbalance-mitigating methods and vanilla mid-concatenation with softmax loss on SSW60 \cite{ssw602022eccv}.}
\label{fig:teaser}
\vspace{-0.6cm}
\end{figure}

Recently, as in-depth investigations of this area, some studies address imbalanced optimization as an obstacle in coarse-grained multi-modal learning \cite{wei2022learning}. The joint multi-modal models optimizing for a uniform objective may have its uni-modal encoders converge at different rates \cite{wang2020makes}, and the potential of multiple modalities can not be fully exploited \cite{peng2022balanced}. To cope with this, extra uni-modal classifiers and auxiliary loss \cite{wang2020makes} are introduced to cultivate better models at the expense of higher training costs. On-the-fly gradient modulation \cite{peng2022balanced} based on approximate uni-modal performance is proposed to enable parameters of different modalities to update with inconsistent learning rates. Some attention-based multi-modal fusion strategies like \cite{nagrani2021attention} allow cross-modal information interaction during training, alleviating the imbalance via information sharing but are highly limited in model architecture. Noting that these studies are set in coarse-grained multi-modal learning, we validate some of them on Audio-Visual Fine-Grained (AVFG) tasks and find they fail on the bird categorization dataset, SSW60 \cite{ssw602022eccv}. From Fig.\ \ref{fig:teaser}, the under-optimized audio modality improves slightly or even gets worse, while the dominant visual modality and the entire model become worse under the regulation of imbalance\footnote{The details of calculating the approximation of uni-modal accuracy in the joint model are in Section \ref{ssec:ana}.}, indicating that the higher demand for the learned feature to be distinguishable in AVFG tasks potentially results in the failure of these methods.

Based on the hypothesis above, we introduce the idea of cosine loss that has shown its power in single-modal fine-grained tasks, such as image classification \cite{liu2016large,liu2017deep}, audio-only speaker verification \cite{wan2018generalized,liu2019speaker}, and face recognition\cite{ranjan2017l2,wang2018cosface, deng2019arcface}, with demonstration by \cite{ranjan2017l2} that a $L_2$ normalization to feature can lower the intra-class angular variability by directly maximizing the margin for the normalized $L_2$ distance or cosine similarity score. Specifically, we propose Multi-Modal Cosine loss, MMCosine, which performs a modality-wise $L_2$ normalization to both features and weight. We further discuss that MMCosine alleviates the imbalance of optimization from two aspects: Handling the problem that the naively trained uni-modal encoder is prone to have a larger weight in norm and coordinating the learning of the angles between weight and features with its intrinsic constraints. 

Compared to previous imbalance-mitigating methods, our MMCosine has little extra training costs and is model or architecture-agnostic. Experiments on AVFG tasks including speaker verification \cite{nagrani2017voxceleb,chung2018voxceleb2}, emotion recognition \cite{cao2014crema} and bird categorization \cite{ssw602022eccv} prove the effectiveness of our method, where MMCosine gains large improvements over vanilla softmax loss. On SSW60, it outperforms other imbalance-mitigating methods. Meanwhile, the simplicity and versatility of our method allow it to stack with other advanced multi-modal fusion strategies to boost the model, indicating the potential of our method to be a paradigm for multi-modal fine-grained learning.
\vspace{-0.4cm}

\section{Method}
\label{sec:pagestyle}
\vspace{-0.2cm}
\subsection{Background}
\label{ssec:subhead}

Mid-concatenation is one of the most common fusion paradigms in multi-modal learning. For this protocol, features from uni-modal encoders are concatenated and go through Fully Connected (FC) layers to get logit scores for each label. The corresponding vanilla softmax loss is presented as:
\begin{equation}\label{formula:1}
\begin{aligned}
L_{vani}&=-\frac{1}{N}\sum_{i=1}^Nlog\frac{e^{W_{y_i}^T[\phi^a_i;\phi^v_i]+b_{y_i}}}{\sum_{j=1}^n e^{W_{j}^T[\phi^a_i;\phi^v_i]+b_{j}}},
\end{aligned}
\end{equation}
where $\phi^a\in \mathbb{R}^d$ and $\phi^v \in \mathbb{R}^d$ denote the features from audio and visual encoders of the $i$-th sample $x_i$ with label $y_i$. The batch size and the class number are $N$ and $n$. $W\in \mathbb{R}^{2d\times n}$ and $b\in{\mathbb{R}^n}$ are weight and bias parameters of FC layers, with $j$ denoting the index of classes. As $W$ can be divided into two modality-wise blocks $[W^a;W^v]$ consistent with $[\phi^a;\phi^v]$, we can rewrite the logit output $f(x_i)_j$ of the $j$-th class of the $i$-th sample $x_i$ as:
\begin{equation}\label{formula:2}
\begin{aligned}
f(x_i)_j={W_j^a}^T\phi_i^a+{W_j^v}^T\phi_i^v+b_j,
\end{aligned}
\end{equation}
then the vanilla softmax loss for mid-concatenated multi-modal learning becomes $L_{vani}=-\frac{1}{N}\sum_{i=1}^Nlog\frac{e^{f(x_i)_{y_i}}}{\sum_{j=1}^n e^{f(x_i)_{j}}}$.

\vspace{-0.2cm}
\subsection{Multi-modal Cosine Loss}
\label{ssec:pipline}

\begin{figure}[t]
\centering
 \centerline{\includegraphics[width=9.5cm]{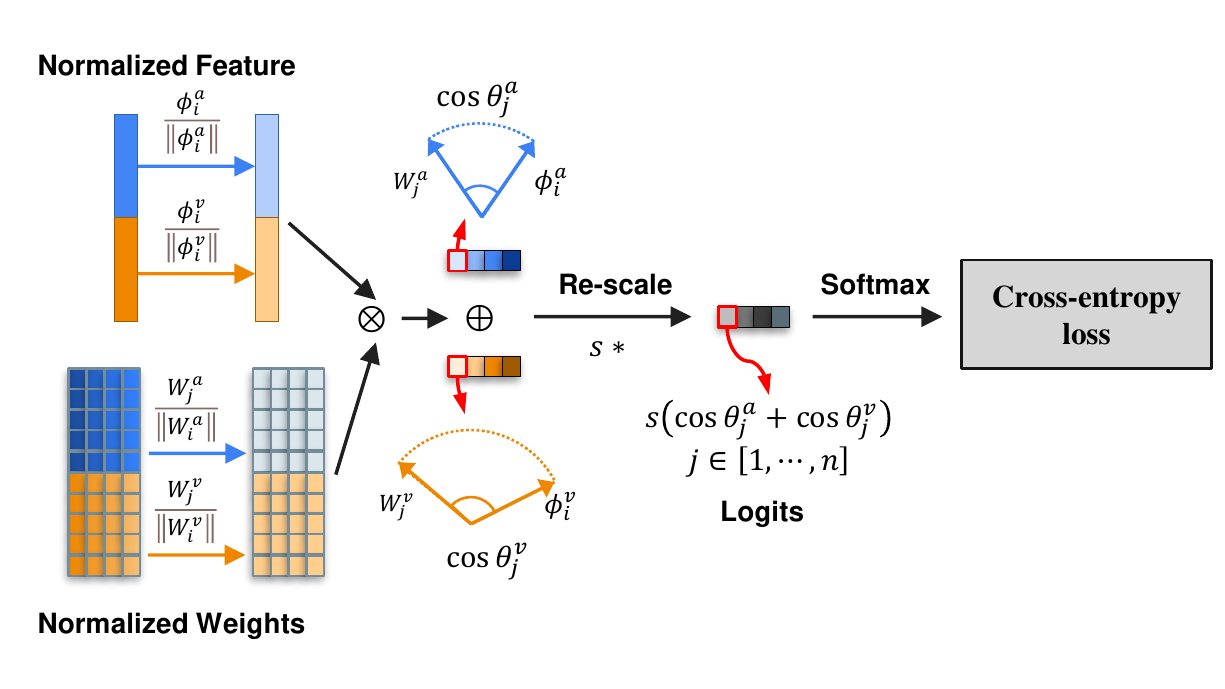}}
  \vspace{-0.6cm}
\caption{The overall pipeline of MMCosine.}
  \vspace{-0.3cm}
\label{fig:pipeline}
\end{figure}

In this subsection, we propose the pipeline of Multi-Modal Cosine loss (MMCosine) to tackle the imbalance problem in AVFG tasks. It alleviates the imbalance from both perspectives of weight norm and its inter-modality constraints, with details in Section \ref{ssec:ana}. As depicted in Fig.\ \ref{fig:pipeline}, we perform a modality-wise $L_2$ normalization to weight and features to replace ${W_j^a}^T\phi_i^a+{W_j^v}^T\phi_i^v$ in Equation (\ref{formula:2}) with $cos\theta_j^a+cos\theta_j^v$, where $cos\theta_j^a=\frac{{W_j^a}^T\phi_i^a}{|| W_j^a||\cdot||\phi_i^a||}$ and similar for $cos\theta_j^v$, with $||\cdot||$ referring to the $L_2$ norm.  Following previous single-modal cosine loss \cite{wang2018cosface,deng2019arcface} in fine-grained learning, the bias term $b$ is omitted for simplicity. We also introduce the scaling parameter $s$ to rescale the combination of uni-modal cosine similarity scores before softmax to guarantee the convergence of the network. In summary, the formal definition of MMCosine can be given as:
\begin{equation}\label{formula:3}
\begin{aligned}
% \tilde{f}(x_i)_j&=\tilde{W}_{j}^a^T\tilde{\phi}^a_i+\tilde{W}_{j}^v^T\tilde{\phi}^v_i=cos\theta^a_j+cos\theta^v_j,\\
L_{mmcosine}% -\frac{1}{N}\sum_{i=1}^Nlog\frac{e^{s\cdot \tilde{f}(x_i)_{y_i}}}{\sum_{j=1}^n e^{s\cdot \tilde{f}(x_i)_{j}}}\\
&=-\frac{1}{N}\sum_{i=1}^Nlog\frac{e^{s\cdot(cos\theta_{y_i}^a+cos\theta_{y_i}^v)}}{\sum_{j=1}^n e^{s\cdot(cos\theta_j^a+cos\theta_j^v)}},
\end{aligned}
\end{equation}
where $\theta_j^a$ is the angle between $W_j^a$ and $\phi^a$ and so is $\theta_j^v$. By MMCosine, we remove the radial variations of weight and features for both uni-modal parts of the joint model. The learned embedding features, $\phi^a$ and $\phi^v$, are expected to distribute on a modality-specific hypersphere, which enables the model to maximize the discriminability of the $L_2$ distance, or cosine similarity, between the modality-wise feature and weight vectors. Following the demonstration in \cite{wang2018cosface}, we also give a lower bound of the scaling parameter $s$ for MMCosine:
\begin{equation}\label{formula:4}
\begin{aligned}
s\geq \frac{C-1}{2(C+1)}log\frac{(C-1)p}{1-p},
\end{aligned}
\end{equation}
where $C$ is the total class number, $p$ is the expected posterior probability for ground-truth. The demonstration and sensitivity analysis can be found in the supplementary material.

\vspace{-0.2cm}
\subsection{Alleviation of Imbalance}

\label{ssec:ana}
\begin{figure}[t]
\begin{minipage}[t]{\linewidth}
  \centerline{}
  {\includegraphics[width=8.8cm]{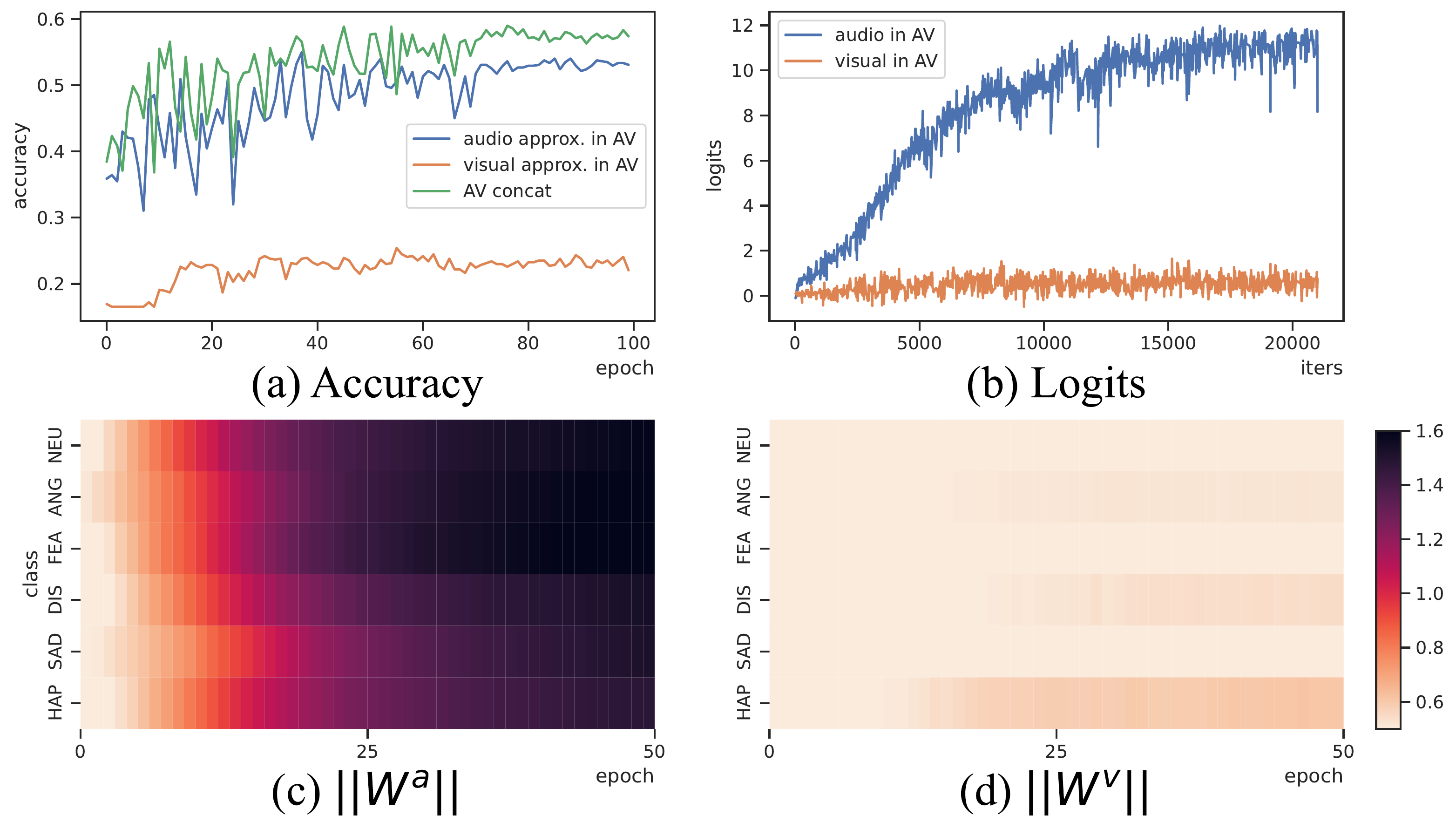}}
\end{minipage}
\vspace{-0.5cm}
\caption{(a) The audio-visual and approximate uni-modal accuracy. (b) The batch-average uni-modal logit scores. (c-d) The observation of modality-wise weight of each label in norm.}
\vspace{-0.5cm}
\label{fig:weight}
\end{figure}

In this subsection, we discuss why MMCosine can close the inter-modality gap of optimization and performance from the perspectives of weight norm and intrinsic constraints.
\\
\textbf{Effect of Weight Normalization.} To explore the limitation of vanilla softmax, we train an audio-visual concatenation-based network on CREMA-D \cite{cao2014crema} while documenting some key indicators. We adopt ${W^m}^T\phi^m+b/2$ to calculate the end-to-end accuracy of modality $m$ \cite{peng2022balanced}, and record the batch-average ${W^m}^T\phi^m$ of the ground-truth as uni-modal logit scores. As shown in Fig.\ \ref{fig:weight}(a) and (b), the dominant audio modality rapidly handles the overall model performance and the joint logit scores, while the visual modality keeps under-optimized throughout the training. By further observation of modality-wise weight in norm of each label, the easily-trained uni-modal encoder tends to have its weight in norm growing much faster than the weak modality, as illustrated in Fig.\ \ref{fig:weight}(c) and (d). The imbalanced modality-wise weight in norm triggers divergent uni-modal logit scores and distorts joint decision-making. 
We also provide a view of gradient propagation to connect weight imbalance with sub-optimization. Denoting the parameters of the audio encoder as $\theta^a$, the gradient of $\theta^a$ deriving from a single sample $x_i$ can be formulated as:
\begin{equation}\label{formula:6}
\begin{aligned}
\frac{\partial L_{vani}}{\partial \theta^a}&=\frac{\partial L_{vani}}{\partial f(x_i)}\cdot \frac{\partial f(x_i)}{\partial \phi^a(\theta^a,x_i)} \cdot \frac{\partial \phi^a(\theta^a,x_i)}{\partial \theta^a},\\
\end{aligned}
\end{equation}
where $\frac{\partial L_{vani}}{\partial f(x_i)_j}=\frac{e^{f(x_i)_j}}{\sum_{j=1}^n e^{f(x_i)_j} }-\mathrm{1}_{j=y_i}$. For one thing, the term of $\frac{\partial L_{vani}}{\partial f(x_i)_{y_i}}$, which is shared by both modalities, approaches zero even if only one of the uni-modality has large weight and logit scores. It indicates that the dominant modality prevents the other from thorough optimization \cite{peng2022balanced}. For another, the $j$-th component of the middle term $\frac{\partial f(x_i)_j}{\partial \phi^a}=\frac{\partial {W_j^a}^T\phi^a}{\partial \phi^a}=W_j^a$. Hence, the uni-modal weight also directly impacts the encoder updating. The imbalanced weight components induce the gradient, and afterward the convergence of uni-modalities to be incoordinate. Therefore, by weight normalization, MMCosine mitigates the imbalance by closing the logit gap and coordinating the optimization process. 
\\
\textbf{Intrinsic Constraints in MMCosine.} According to Equation (\ref{formula:3}), the logit score is the combination of cosine similarity scores. By naive trigonometric transformation, the logit score $\tilde{f}(x_i)_j$ for class $j$ can be written as follows:
\vspace{-0.5em}\begin{equation}\label{formula:5}
\begin{aligned}
\tilde{f}(x_i)_j=cos\theta^a_j+cos\theta^v_j=2cos(\frac{\theta^a_j+\theta^v_j}{2})\cdot cos(\frac{\theta^a_j-\theta^v_j}{2}).
\end{aligned}
\end{equation}
This formula provides a simple and informative view that MMCosine has symmetric constraints on the cooperation and discrepancy of multiple modalities. Considering $\theta^a_j,\theta^v_j$ as a surrogate in the modality-wise hypersphere for uni-modal confidence for class $j$, the logit score approaches a large value only when both modalities have high confidence, \textit{i.e.}, $\theta^a_j$ and $\theta^v_j$ are both small. It can be interpreted as a constraint on cooperation. Meanwhile, $\theta^a_j-\theta^v_j$ also has to be small, or $\theta^a_j$ and $\theta^v_j$ have to keep close with the symmetric constraint, which requires a more balanced optimization of uni-modalities. Generally, the intrinsic constraints in MMCosine originate from the removed radial variation of weight and features and the bounded domain of value for logit scores. In contrast, vanilla softmax allows high logit scores even if only one modality is fully trained with the infinite range of ${W_j^m}^T\phi^m$ ($m$ refers to the index of the uni-modality). Therefore, MMCosine encourages a more balanced and effective multi-modal learning than vanilla softmax loss with its intrinsic constraints.
% \begin{equation}
% \begin{aligned}
% \rho_{av}=\frac{||\frac{\partial f(x_i)_j}{\partial \phi^a}||}{||\frac{\partial f(x_i)_j}{\partial \phi^v}||}=\frac{||W^a_j||}{||W^v_j||}
% \end{aligned}
% \end{equation}

% \begin{figure}[t]
% %
% \begin{minipage}[t]{.25\linewidth}
%   \centering
%   \centerline{\includegraphics[width=2.5cm]{vanilla_acc.pdf}}
% %  \vspace{-0.1cm}
%   \centerline{(a) Accuracy }\medskip
% \end{minipage}
% \hfill
% \begin{minipage}[t]{.24\linewidth}
%   \centering
%   \centerline{\includegraphics[width=2.5cm]{vanilla_logit.pdf}}
% %   \vspace{-0.1cm}
% %  \vspace{1.5cm}
%   \centerline{(b) Logit }\medskip
% \end{minipage}
% % \vspace{-0.5cm}
% \hfill
% \begin{minipage}[t]{.24\linewidth}
%   \centering
%   \centerline{\includegraphics[width=2.5cm]{weight_a.pdf}}
% %  \vspace{1.5cm}
% % \vspace{-0.1cm}
%   \centerline{(c) $W^a$ norm }\medskip
% \end{minipage}
% \hfill
% \begin{minipage}[t]{.24\linewidth}
%   \centering
%   \centerline{\includegraphics[width=2.5cm]{weight_v.pdf}}
% %   \vspace{-0.1cm}
% %  \vspace{1.5cm}
%   \centerline{(d) $W^v$ norm}\medskip
% \end{minipage}
% \vspace{-0.4cm}
% \caption{(a) The audio-visual and approximate uni-modal accuracy. (b) The batch-average uni-modal logit scores. (c,d) The observation of modality-wise weight of each label in norm.}
% \label{fig:weight}
% % \end{figure}

\begin{table*}[t]
\begin{center}
\begin{tabularx}{\linewidth}{cccccccccc}
\toprule

\multicolumn{2}{c|}{\multirow{2}{*}{\textbf{Method}}} & \multicolumn{2}{c}{\textbf{CREMA-D}} &\multicolumn{2}{c}{\textbf{SSW60}} & \multicolumn{4}{c}{\textbf{Voxceleb}}\\

\cmidrule(lr){3-4}\cmidrule(lr){5-6}\cmidrule(lr){7-10}

\multicolumn{2}{c|}{} &\multicolumn{2}{|c}{Top1-Accuracy($\%$)} & \multicolumn{2}{c}{Top1-Accuracy($\%$)} & VC1 EER($\%$) & VC1 minDCF & VC2 EER($\%$) & VC2 minDCF  \\
\midrule

\multicolumn{2}{c|}{Mid-concat}  &\multicolumn{2}{|c}{60.08}	&\multicolumn{2}{c}{73.32} &6.81	&0.578	&6.21	&0.580\\ 
% \hline

% \multicolumn{2}{|c|}{late fusion}  &\multicolumn{2}{|c|}{61.02}	&\multicolumn{2}{|c|}{73.07} & 6.17	&0.596&	6.36& 0.549\\ 
% \hline

\multicolumn{2}{c|}{FiLM}  &\multicolumn{2}{|c}{59.68}	&\multicolumn{2}{c}{71.67} &11.50	&0.537	&8.31	&0.644\\ 
% \hline

\multicolumn{2}{c|}{Gated}  &\multicolumn{2}{|c}{60.48}	&\multicolumn{2}{c}{70.64} &10.39	&0.567	&7.76	&0.640\\ 
\midrule

\multicolumn{2}{c|}{Mid-concat$\dagger$}  &\multicolumn{2}{|c}{63.44}	&\multicolumn{2}{c}{\textbf{75.95}} &\textbf{4.26}	&0.461	&\textbf{4.13}	&0.371\\ 
% \hline

\multicolumn{2}{c|}{FiLM$\dagger$}  &\multicolumn{2}{|c}{61.42}	&\multicolumn{2}{c}{74.30} &8.03	& 0.373	&4.58	&0.342\\ 
% \hline

\multicolumn{2}{c|}{Gated$\dagger$}  &\multicolumn{2}{|c}{\textbf{66.40}}	&\multicolumn{2}{c}{75.70} &5.34	&\textbf{0.335}	&4.30	&\textbf{0.322}\\ 
\bottomrule

\end{tabularx}

\end{center}
\vspace{-1.5em}
\caption{Performance of various fusion strategies on three AVFG tasks. $\dagger$ indicates MMCosine is applied. Combined with MMCosine, most of the fusion methods gain considerable improvement. }\label{table_main}
\vspace{-0.38cm}
\end{table*}

% 缩减数据集介绍
% 模型建立
% 讨论分条
\vspace{-0.3cm}
\section{EXPERIMENTS}
\label{sec:typestyle}
\vspace{-0.1cm}
\subsection{Audio-visual Fine-grained Tasks and Model Setup}
\label{ssec:dataset}

% \\VoxCeleb1 consists of more than 150,000 utterances from 1251 celebrities, and VoxCeleb2 consists of more than 1,000,000 utterances from 6112 celebrities.

\textbf{Speaker Verification.} Voxceleb provides roughly 1M utterances from 7K speakers for speaker verification. Following previous study \cite{chen2020multi,xie2019utterance,sari2021multi}, we train our model on Voxceleb2 \cite{chung2018voxceleb2} dev partition and validate on Voxceleb1 \cite{nagrani2017voxceleb} and the trial list randomly sampled from Voxceleb2 test. A part of the video links of Voxceleb1 test are unavailable. We use Retinaface \cite{deng2020retinaface} to extract and align faces. Equal Error Rate (EER) and minimum Detection Cost Function (mDCF) are employed as the performance metrics.
\\
\textbf{Emotion Recognition.} CREMA-D \cite{cao2014crema} is an audio-visual dataset utilizing face and voice data to classify six basic emotional states and contains 7,442 video clips and is divided into training and validation sets at a ratio of 9 to 1.
\\
\textbf{Bird Categorization.} The up-to-date fine-grained dataset SSW60 \cite{ssw602022eccv} for bird categorization provides 5,400 video clips of 60 species of birds and additional unpaired images and audio segments to pretrain the uni-modal encoder. Joint models initialized with the same pretrained uni-modal backbones are trained on video clips with MMCosine and other methods.
\\
\textbf{Model Setup.}  For Voxceleb/CREMA-D/SSW60, we employ ResNet-18/18/50 as uni-modal backbones. The audio and visual branches of each task have the same feature dimension of 512/512/1024. The channel of audio ResNet is reduced to 1 to fit the sound spectrogram with sample rate of 16kHz. For each task, the scaling parameter $s$ is roughly set by Equation (\ref{formula:4}) without elaborate searching. All the experiments are conducted on four NVIDIA 3090Ti GPUs.
\vspace{-0.2cm}
\subsection{Versatility}
\label{ssec:subhead2}

As MMCosine is proposed in the mid-concatenation base, we attempt to verify its versatility with other advanced two-stream fusion strategies: FiLM \cite{perez2018film} and Gated \cite{kiela2018efficient}. We adapt these non-symmetric fusion methods by adding a stream where the dual modality features switch their roles. Meanwhile, as MMCosine is an orthogonal exploration to previous imbalance-mitigating methods, we first compare MMCosine with OGM-GE \cite{peng2022balanced} and G-blending \cite{wang2020makes} to verify that MMCosine suits better in fine-grained tasks, and then stack them up to investigate whether MMCosine can further boost these methods. For G-blending, we adopt MMCosine in the multi-modal head and simple cosine loss with equal scaling factors in the extra uni-modal heads.

\begin{table}[t]

\begin{center}
\begin{tabularx}{0.8\linewidth}{c|cc}
\toprule
\textbf{ Metric}&\textbf{Vanilla Softmax} &\textbf{MMCosine}\\

\midrule
A-probe &56.32&48.66\\
V-probe &32.26&42.40\\
\midrule
A-V gap& 24.06 &6.26\\
% \midrule
% \multirow{2}{*}{SSW60}&A-probe &20.54&21.10\\
% {}&V-probe &70.69&72.50\\
\bottomrule
\end{tabularx}
\end{center}
\vspace{-1.5em}
\caption{Uni-modal accuracy by linear-probe on CREMA-D. }\label{table_probe}
\vspace{-1.5em}
\end{table}

\vspace{-0.3cm}
\subsection{Results and Discussion}
\begin{figure}[t]
\begin{minipage}[b]{\linewidth}
  \centering
  \centerline{\includegraphics[width=8.8cm]{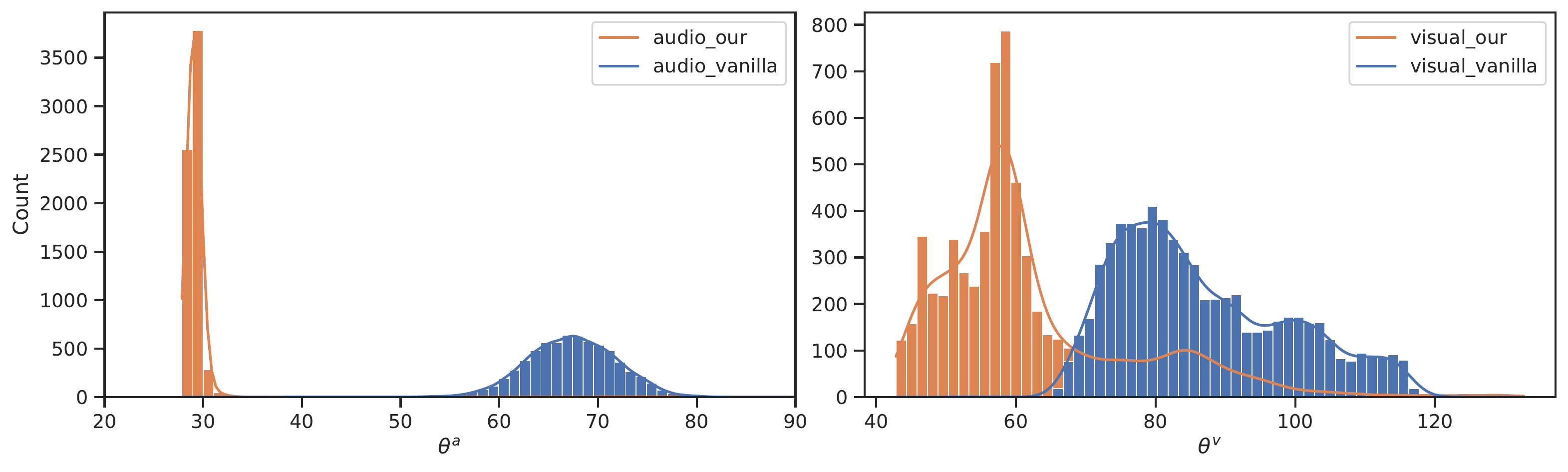}}
%  \vspace{1.5cm}
  \vspace{-1.5em}
\end{minipage}
\caption{ Distribution of learned uni-modal angles. Left: audio; Right: visual. Orange: MMCosine; Blue: Vanilla softmax. }
\vspace{-0.15cm}
\label{fig:theta}
\end{figure}

\vspace{-0.1cm}
\label{ssec:subhead1}
\textbf{Performance Enhancement\footnote{See supplementary for additional results on coarse-grained datasets.}.} As shown in Tab.\ \ref{table_main}, MMCosine outperforms vanilla softmax by considerable margins on almost all the fusion methods. The performance promotion on fine-grained datasets of various scales, domains, and label amounts indicates that MMCosine can universally achieve a more distinguishable feature learning. 
\\
\textbf{Imbalance Mitigation.} For our main concern of imbalance issue, we use linear-probe \cite{alain2016understanding} to detect the quality of cultivated uni-modal backbones accurately. From Tab.\ \ref{table_probe}, the performance gap of uni-modal encoders is reduced by MMCosine, with the weak modality and the joint model boosted. It proves that MMCosine can balance the optimization towards better joint learning.
\\
\textbf{Lower Angular Variance.} 
Fig.\ \ref{fig:theta} illustrates the learned angles between uni-modal features and ground-truth weight vectors, or class centers. Both uni-modal angle distributions in MMCosine shift towards smaller value and become more compact. It indicates that MMCosine can lower the intra-class angular variation and maximize the discriminability of cosine metric, which suits fine-grained tasks better.
\\
\textbf{Versatility.} From Tab.\ \ref{table_main}, FiLM and Gated are mostly inferior to simple concatenation. It makes sense as the mixed uni-modal information further distorts the already non-convex decision boundary. However, these fusion methods all gain significant improvement with MMCosine and beat concatenation in some cases, suggesting that MMCosine can enhance the generalization of these methods to harder tasks.  
\\
\textbf{Experiments with Other Imbalance-mitigating Methods.} From Tab.\ \ref{table_balance}, MMCosine outperforms OGM-GE and G-blending on SSW60. It shows the superiority of MMCosine on feature separability. It should also be noted that MMCosine is more cost-effective and easy-to-use.  When stacked up with OGM-GE and G-blending, MMCosine, in turn, helps them learn better, which reveals the potential of MMCosine as a paradigm for AVFG tasks.

\begin{table}[t]
\begin{center}
\begin{tabularx}{0.8\linewidth}{ccc}
\toprule
\textbf{Method}&\textbf{Vanilla Softmax}&\textbf{MMCosine}\\
\midrule
Mid-concat& 73.32&75.95\\
OGM-GE & 72.50&74.30 \\
G-blending &72.24&74.51\\
\bottomrule
\end{tabularx}
\end{center}
\vspace{-1.5em}
\caption{Results of MMCosine compared with other imbalance-mitigating methods on SSW60. }\label{table_balance}
\vspace{-1.5em}
\end{table}

\vspace{-0.4cm}
\section{CONCLUSION}
\label{sec:majhead}
\vspace{-0.2cm}
In this work, we present Multi-Modal Cosine Loss, MMCosine, to alleviate the imbalance issue and boost model performance in audio-visual fine-grained learning. Extensive experiments on various tasks prove the effectiveness of our method. The simplicity and high versatility of MMCosine make it potentially serve as a paradigm. For future work, it would also be worthwhile to generalize MMCosine to coarse-grained tasks and scenarios with triple or more modalities. 

\vspace{-0.4cm}
\section{Acknowledgements}
\vspace{-0.2cm}
This research was supported by National Natural Science Foundation of China (NO.62106272), the Young Elite Scientists Sponsorship Program by CAST (2021QNRC001), and Public Computing Cloud, Renmin University of China.
\newpage
\small
\bibliographystyle{IEEEbib}
\bibliography{main}

\end{document}